\documentclass[12pt]{iopart}

\usepackage{graphicx}

 \def\gsim{\mathrel{\rlap{\lower4pt\hbox{\hskip1pt$\sim$}}
 \raise1pt\hbox{$>$}}}

 \newcommand\beq{\begin{equation}}
 
 \newcommand\eeq{\end{equation}}
 \newcommand\beqn{\begin{eqnarray}}
 \newcommand\eeqn{\end{eqnarray}}
\def\mb{\,\mbox{mb}}
\def\mub{\,\mbox{$\mu$b}}

\def\GeV{\,\mbox{GeV}}

\def\lsim{\mathrel{\rlap{\lower4pt\hbox{\hskip1pt$\sim$}}
    \raise1pt\hbox{$<$}}}         %less than or approx. symbol
\def\gsim{\mathrel{\rlap{\lower4pt\hbox{\hskip1pt$\sim$}}
    \raise1pt\hbox{$>$}}}         %greater than or approx. symbol

\def\mb{\,\mbox{mb}}

\def\GeV{\,\mbox{GeV}}

\begin{document}

\title{Vector meson production in ultra-peripheral
collisions at LHC}

\author{Yuri Ivanov$^{1,2}$, Boris Kopeliovich$^{1-3}$ and Ivan
Schmidt$^{1}$}

\address{$^1$Departamento de F\'{\i}sica y Centro de Estudios
Subat\'omicos,
Universidad T\'ecnica Federico Santa Mar\'{\i}a,
Casilla 110-V,
Valpara\'\i so, Chile}
\address{$^{2}$
Joint Institute
for Nuclear Research, Dubna, Russia}
\address{$^{3}$
Institute for Theoretical Physics, University of Heidelberg,
Germany }

 \begin{abstract}
 Vector meson production in ultra-peripheral collisions of heavy ions at
LHC is calculated, in VDM and color dipole approaches.
 \end{abstract}

%\section{UPC}

One of the intensive channels in ultra-peripheral collisions of
heavy ions (UPC) is photoproduction of vector mesons. If the
nucleus remains in the ground state, $\gamma A\to VA$, such
reaction is called coherent, otherwise we call it incoherent.

\begin{figure}[h]
\centerline{
  \scalebox{0.63}{\includegraphics{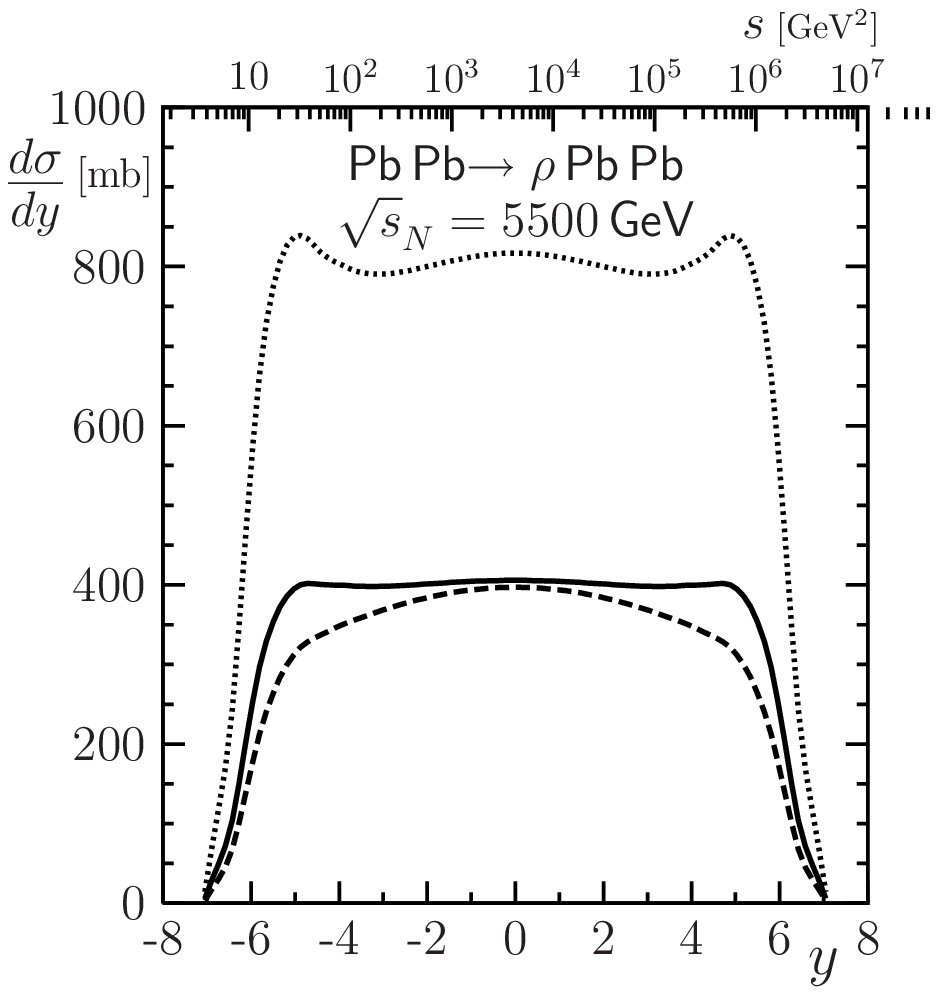}}
  \scalebox{0.63}{\includegraphics{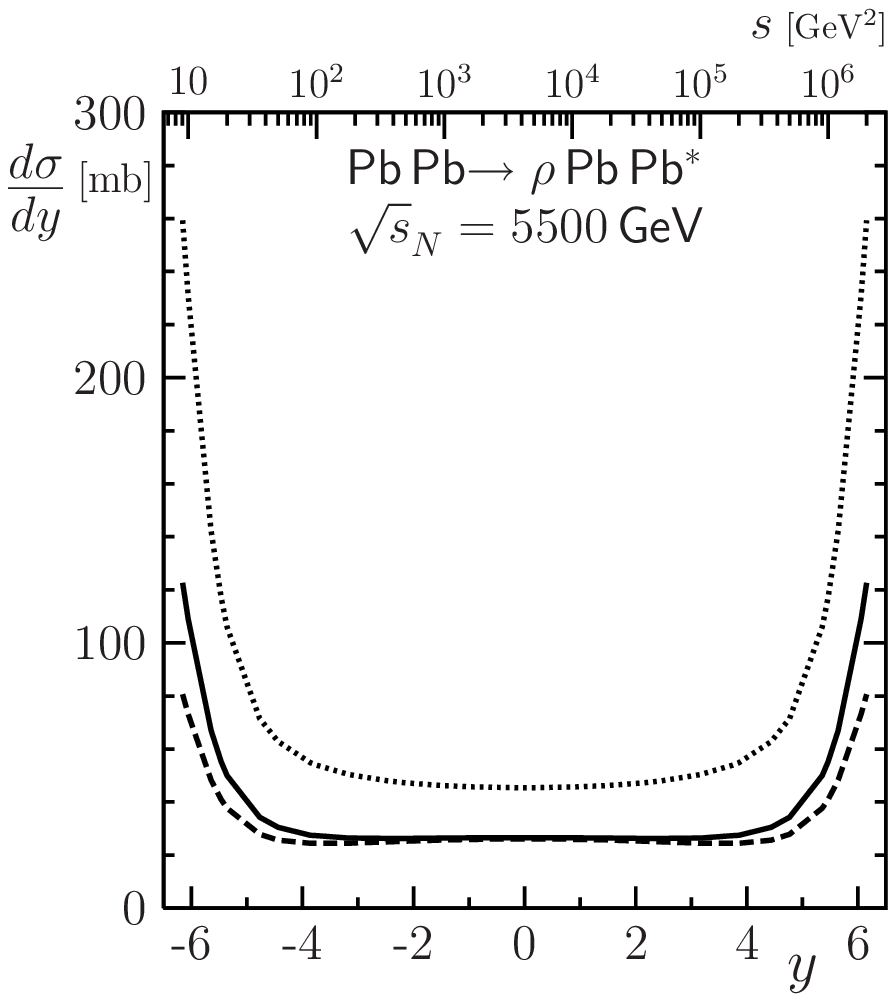}}
 }
\caption{\label{fig:rho}\em
 Rapidity distributions of coherent (left) and incoherent (right) $\rho$
photoproduction, calculated both in VDM (dotted line) and color
dipole approaches.  Dashed and solid curves correspond to the KST
and Reggeon corrected parametrizations of the dipole cross
section.}
 \vspace*{-3cm}
 \end{figure}

A simple approach for calculating the cross sections is the
vector-dominance model (VDM) combined with the Glauber eikonal
approximation \cite{glauber}. The results of such calculation are
depicted in fig.~\ref{fig:rho} by dotted curves.

Gribov corrections are calculated within the color dipole approach
\cite{hikt}, using the saturated shape (KST parametrization
\cite{kst2}) of the dipole cross section adjusted to low $Q^2$
data. The yield of $\rho$-mesons is considerably reduced.

\begin{figure}[h]
 \vspace*{-1.5cm}
\centerline{
  \scalebox{0.63}{\includegraphics{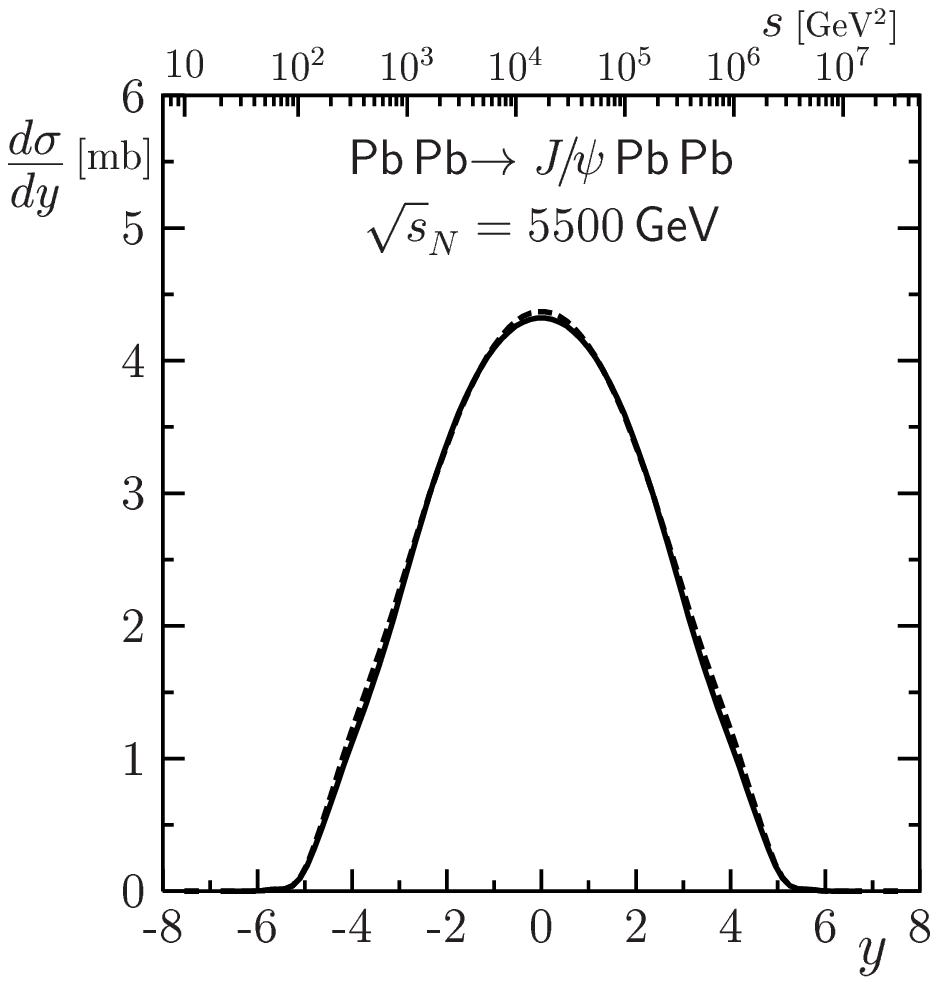}}
  \scalebox{0.63}{\includegraphics{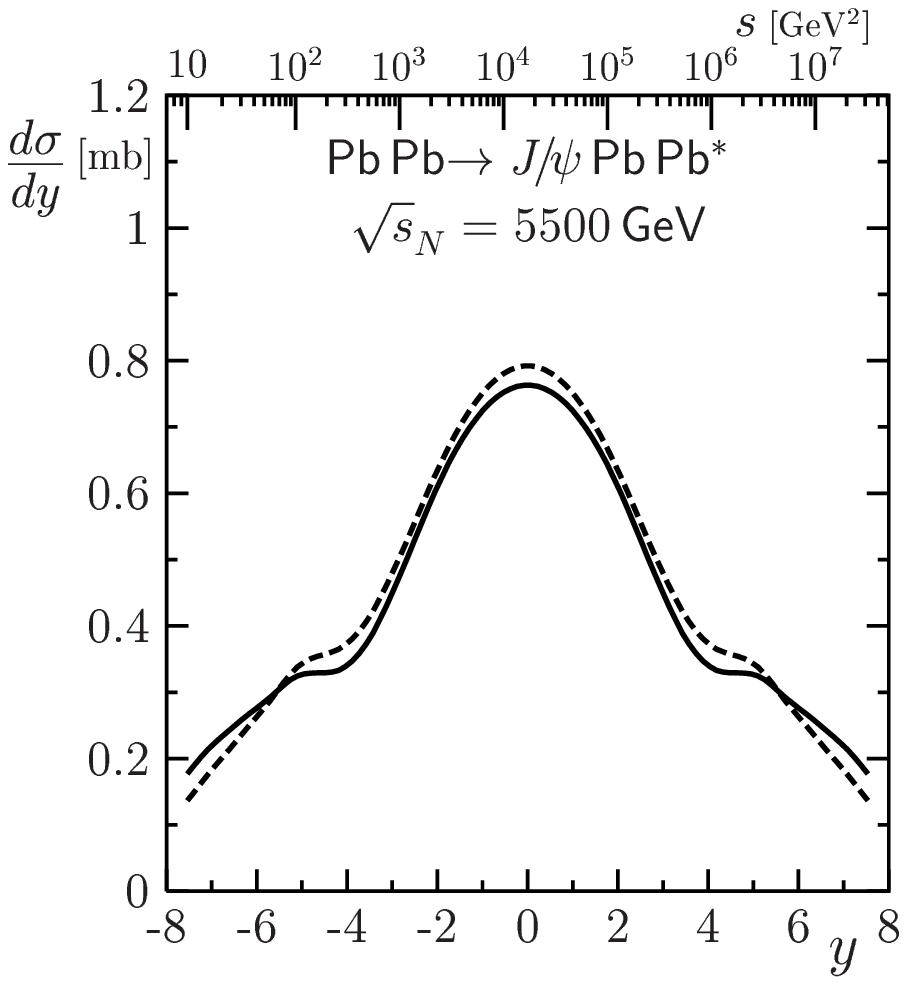}}
 }
\caption{\label{fig:psi}\em
 Rapidity distributions of coherent (left) and incoherent (right) $J/Psi$
photoproduction, at the energy of LHC. Solid and dashed curves
correspond to the KST \cite{kst2} and GBW \cite{gbw}
parametrizations of the dipole cross section.}
  \end{figure}

Photoproduction of charmonia can be calculated more reliably
\cite{hikt}, since the wave function of a nonrelativistic $\bar
cc$ system is known much better. Fig.~\ref{fig:psi} presents the
results for coherent and incoherent photoproduction of $J/\Psi$ in
lead-lead collisions at LHC. Calculations were performed with two
parametrizations for the dipole cross section, KST \cite{kst2} and
GBW \cite{gbw}, fitted to low and high $Q^2$ data. The difference
is small.

The rapidity integrated cross sections for $\rho$ and $J/\Psi$
production, and at RHIC and LHC energies, are shown in table~1.
 \begin{table}[h]
\centerline{
\begin{tabular}{|l|c|c|c|}
 \hline
 & $\sqrt{s}=130\GeV$
 & $\sqrt{s}=200\GeV$
 & $\sqrt{s}=5500\GeV$\\
 \hline
 $J/\Psi$: KST & $71.6\mub$ & $274\mub$ & $26.3\mb$\\
 $J/\Psi$: GBW & $78.3\mub$ & $304\mub$ & $26.7\mb$\\
 \hline
 $\rho$: KST-R & $380\mb$  & $605\mb$  & ~$4.90\,$barn\\
 $\rho$: KST   & $296\mb$  & $483\mb$  & ~$4.36\,$barn\\
 $\rho$: GBW   & $293\mb$  & $478\mb$  & ~$3.99\,$barn\\
 $\rho$: VDM   & $812\mb$  & $1278\mb$ & $10.03\,$barn\\
 \hline
\end{tabular}
}
\caption{
 Total cross sections for coherent production of $\rho$ and $J/\Psi$. }
 \end{table}

 \vspace*{3cm}

\ack This work was supported in part by Fondecyt (Chile) grant 1050519 and
by DFG (Germany)  grant PI182/3-1.


\section*{References}
\begin{thebibliography}{10}

\bibitem{glauber} H\"ufner J, Kopeliovich B Z and Nemchik J
1996 \PL B{\bf 383} 362

\bibitem{hikt} H\"ufner J, Ivanov Yu P, Kopeliovich B Z and Tarasov A V
2000 \PR D{\bf 62} 094022

\bibitem{kst2} Kopeliovich B Z, Sch\"afer A and
Tarasov A V 2000 \PR D{\bf 62} 054022

\bibitem{gbw}
Golec-Biernat K and Wusthoff M, \PR 1999 D{\bf 59} 014017

\end{thebibliography}
 \end{document}